\documentclass[jkps,preprint,fleqn,showpacs,showkeys]{revtex4}
\usepackage{graphicx}
\usepackage{amssymb}
\usepackage{amsmath}
\usepackage{epstopdf}
\usepackage{bm}
\begin{document}
\setcounter{page}{1}
\title{Generalized analytical solutions for secure transmission of signals using a simple communication scheme with numerical and experimental confirmation}
\author{G. Sivaganesh}
\affiliation{Department of Physics, Alagappa Chettiar College of Engineering $\&$ Technology, Karaikudi, Tamilnadu-630 004, India}
\author{A. Arulgnanam}
\email{gospelin@gmail.com}
\affiliation{Department of Physics, St. John's College, Palayamkottai, Tamilnadu-627 002, India}
\author{A. N. Seethalakshmi}
\affiliation{Department of Physics, The M.D.T Hindu College, Tirunelveli, Tamilnadu - 627 010, India}


\begin{abstract}
A novel explicit analytical solution is reported for the transmission and recovery of information signals using a simple communication scheme. Analytical solutions are obtained for the normalized state equations of coupled second-order chaotic transmitter and receiver systems embedding the information signal. The analytical solution of the difference system obtained from the state equations of the transmitter and receiver systems has been identified as a measure of the recovered information signal which is transmitted securely by chaotic masking. The analytical solutions are used to reveal the nature of synchronization and the enhancement of the amplitude of recovered information signal. The difference signal of the coupled state variables indicating the recovered information signal obtained through numerical simulations is presented to validate the analytical results. The electronic circuit experimental results are presented to confirm the analytical and numerical results of the communication scheme discussed.
\end{abstract}

\pacs{05.45.Xt, 05.45.-a}

\keywords{chaos; synchronization; signal transmission}

\maketitle

\section{INTRODUCTION}
\label{sec:1}

The concept of secure communication using chaotic carrier signals has its origin in the phenomenon of chaos synchronization. Following the {\emph{master-slave}} concept of chaos synchronization introduced by Pecora and Carroll \cite{Pecora1990}, a variety of complex dynamical systems have been studied for synchronization \cite{Rulkov1992,Chua1992,Chua1993,Murali1993,Boccaletti2002}. The method proposed by Pecora {\emph{et al.}} involves the synchronization of a {\emph{slave}} sub-system with a {\emph{drive}} subsystem. The synchronization phenomena observed in electronic circuit systems has been used for the secure transmission of information signals through chaotic masking i.e. {\emph{spread spectrum communications}} \cite{Chua1992,Murali1993,Oppenheim1992,Murali1994,Murali1997,Koronovskii2009} and for signal transmission through image encryption \cite{Murali2001,Murali2003,Wu2019}. Further, several electronic circuit systems with chaotic and hyperchaotic behavior have been designed and analyzed for the implementation of secure transmission of signals \cite{Wang2019,Kocamaz2018,Roy2018,Cicek2018,Ren2017,Liu2016,Saini2014}. Synchronization is also observed in coupled chaotic systems by unidirectional (one-way) coupling of the master and slave systems using a common chaotic signal without constructing any stable subsystems. This method has been successfully implemented to achieve synchronization and for signal transmission by chaotic masking of the information signal \cite{Murali1993,Murali1994}. The identification of chaotic dynamics in a second-order, non-autonomous chaotic circuit by {\emph{Murali et al.}} \cite{Murali1994a} has opened the gateway for designing several simple circuit systems with high complex dynamics \cite{Thamilmaran2000,Arulgnanam2009,Arulgnanam2015}. The piecewise-linear nature of the nonlinear elements present in these simple systems makes their dynamics to be mathematically tractable. The dynamical process of chaos synchronization observed in unidirectionally and mutually coupled simple chaotic systems have been greatly studied through explicit analytical solutions, numerical simulations and confirmed experimentally \cite{Sivaganesh2015,Sivaganesh2017,Sivaganesh2018,Sivaganesh2018b}. A mathematical analysis based on the time series of the state variables has been presented for the synchronization and signal transmission using chaos in the quadratic and {\emph{Ueda}} systems \cite{Jovic2006}. However, an explicit analytical solution for the normalized state equations of the coupled transmitter and receiver systems explaining the transmission and recovery of information signals using simple chaotic systems is yet to be studied. The mathematical tractable nature of the systems may provide new insights on the synchronization dynamics of the systems upon signal transmission.\\

This paper is aimed at presenting a novel generalized explicit analytical solution for the transmission and recovery of information signals using the simple communication scheme proposed by {\emph{Murali et al.}} \cite{Murali1994} using the chaotic carrier signals obtained from a class of simple second-order chaotic systems. The forced series {\emph{LCR}} circuit system with two types of piecewise-linear elements namely, the {\emph{Chua's diode}} and the {\emph{simplified nonlinear element (SNE)}}, are studied for the transmission of information. The nature of the synchronization phenomena observed in the coupled systems upon the recovery of information signal is studied analytically. The analytical results thus obtained are validated through numerical and experimental results. This paper is divided as follows. In Sec. \ref{sec:2} we introduce a simple communication scheme and present the experimental results for the transmission of a signal by chaotic masking observed in coupled simple chaotic systems. The analytical and numerical results for signal transmission is presented in Sec. \ref{sec:3} and Sec. \ref{sec:4}, respectively. 

\section{Experimental results}
\label{sec:2}
\begin{figure}
\begin{center}
\includegraphics[scale=0.6]{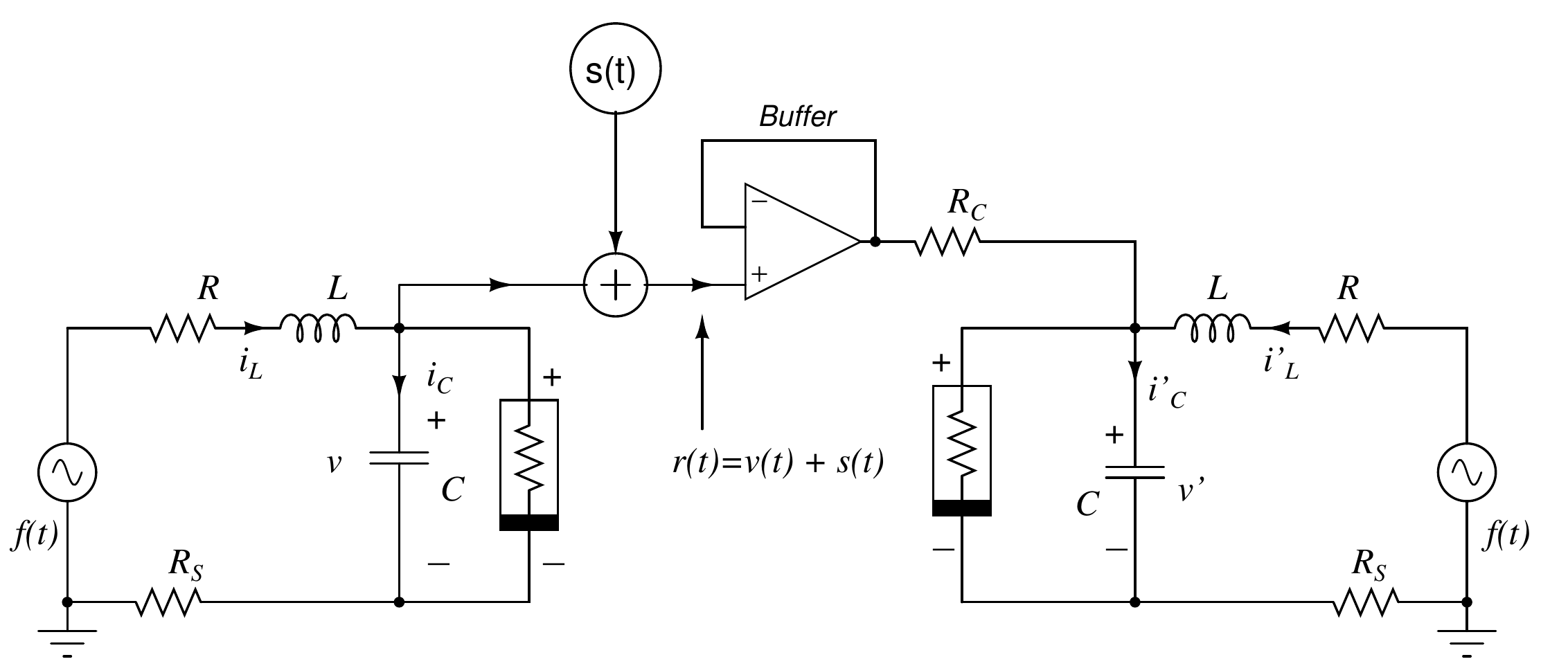}
\caption{Circuit realization of two forced series {\emph{LCR}} circuits with unidirectional coupling for signal transmission.}
\label{fig:1}
\end{center}
\end{figure}

In this section, we present the experimental results on the design of a unidirectionally coupled transmitter and receiver to illustrate the retrieval of the information signal. The applicability of this method of chaos synchronization and signal masking approach to coupled non-autonomous series {\emph{LCR}} circuit systems under the {\emph{drive-response}} configuration is discussed. The physical implementation of the circuitry for chaos communication in a forced series {\emph{LCR}} circuit is as shown in Fig. \ref{fig:1}. The circuits on the left and the right side indicate the transmitter and receiver systems, respectively. The circuit equations governing the transmitter and receiver system with the information signal as shown in Fig. \ref{fig:1} can be written as\\

\begin{figure}
\begin{center}
\includegraphics[scale=0.6]{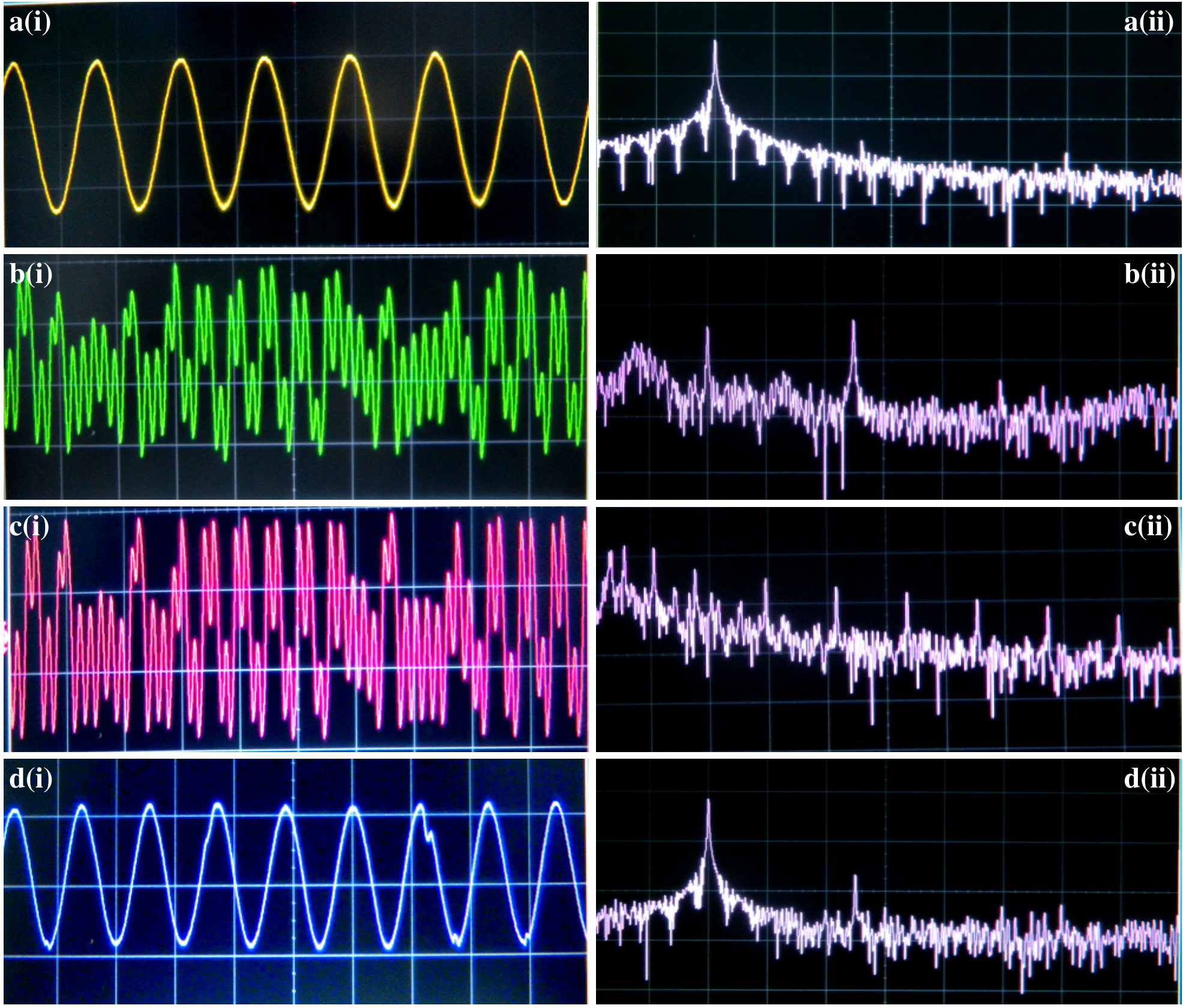}
\caption{Experimental implementation of the communication scheme observed using the unidirectionally coupled forced series {\emph{LCR}} circuits indicating the (i) time series and the corresponding (ii) power spectrum of the signals; (a) sine wave information signal ($s(t)= F sin \omega t$, $F=0.5$ V, $\nu=6280$ Hz; (b) chaotic carrier signal $v(t)$; (c) the amplitude modulated chaotic transmitted signal $r(t) = v(t)+s(t)$ and (d) the recovered information signal $s^{'}(t)$ at the receiver.}
\label{fig:2}
\end{center}
\end{figure}

{\emph{Drive}} - Transmitter:
\begin{subequations}
\begin{eqnarray} 
C {dv \over dt } & = & i_L - g(v), \\
L {di_L \over dt } & = & -R i_L - R_s i_L - v + F_1 sin( \Omega_1 t),
\end{eqnarray}
\label{eqn:1}
\end{subequations}
{\emph{Response}} - Receiver:
\begin{subequations}
\begin{eqnarray} 
C {dv^{'} \over dt } & = & i^{'}_L - g(v^{'}) + \epsilon (r(t)-v^{'}), \\
L {di^{'}_L \over dt } & = & -R i^{'}_L - R_s i^{'}_L - v^{'} + F_2 sin( \Omega_2 t),
\end{eqnarray}
\label{eqn:2}
\end{subequations}
where $g(v)$ and $g(v^{'})$ representing the piecewise linear element given by
\begin{subequations}
\begin{eqnarray} 
g(v) &=& G_b v + 0.5(G_b - G_a)[|v+B_p|-|v-B_p|],\\
g(v^{'}) &=& G_b v^{'} + 0.5(G_b - G_a)[|v^{'}+B_p|-|v^{'}-B_p|],
\end{eqnarray}
\label{eqn:3}
\end{subequations}
The terms $\epsilon = \frac{R}{R_c}$ and $r(t)=v(t)+s(t)$ where, $\epsilon$ and $s(t)$ represent the coupling strength and the transmitted signal, respectively. The drive system operating at the double-band chaotic attractor state is used as the carrier wave to mask the information signal. The typical experimental information signal {\emph{s(t)}} is shown in Fig. \ref{fig:2}a(i) and the experimentally observed double-band chaotic waveform, {\emph{v(t)}} observed across the capacitor $C$ is shown in Fig. \ref{fig:2}b(i). The chaotic signal {\emph{v(t)}} has been used as the carrier wave of this unidirectional scheme. In the absence of the signal {\emph{s(t)}}, a synchronized chaotic behavior is observed in the coupled system between {\emph{v(t)}} and $v^{'}(t)$ \cite{Sivaganesh2018}, for $\epsilon=1105.5$, $C=10.31$ nF, $L=51.3$ mH, $R=2211~\Omega$, $F_{1,2}=1.7$ V, $\nu_{1,2}=5500$ Hz, $G_a= -0.56$ mS, $G_b = 2.5$ mS and $B_p = \pm3.8$ V. When the power of the information signal {\emph{s(t)}} is comparably lower than the chaotic carrier signal {\emph{v(t)}} then {\emph{s(t)}} can be recovered as \cite{Lakshmanan1995a}
\begin{equation}
s(t) = r(t) - v^{'}(t) = v(t)+s(t) - v^{'}(t) \approx s^{'}(t),
\label{eqn:4}
\end{equation}
where $s^{'}(t)$ represents the information signal recovered at the receiver system. Experimental observation of Fig. \ref{fig:2} gives rise to the information signal $s^{'}(t)$ as recovered at the response system by adopting Eq. \ref{eqn:4}.\\
\begin{figure}
\begin{center}
\includegraphics[scale=0.6]{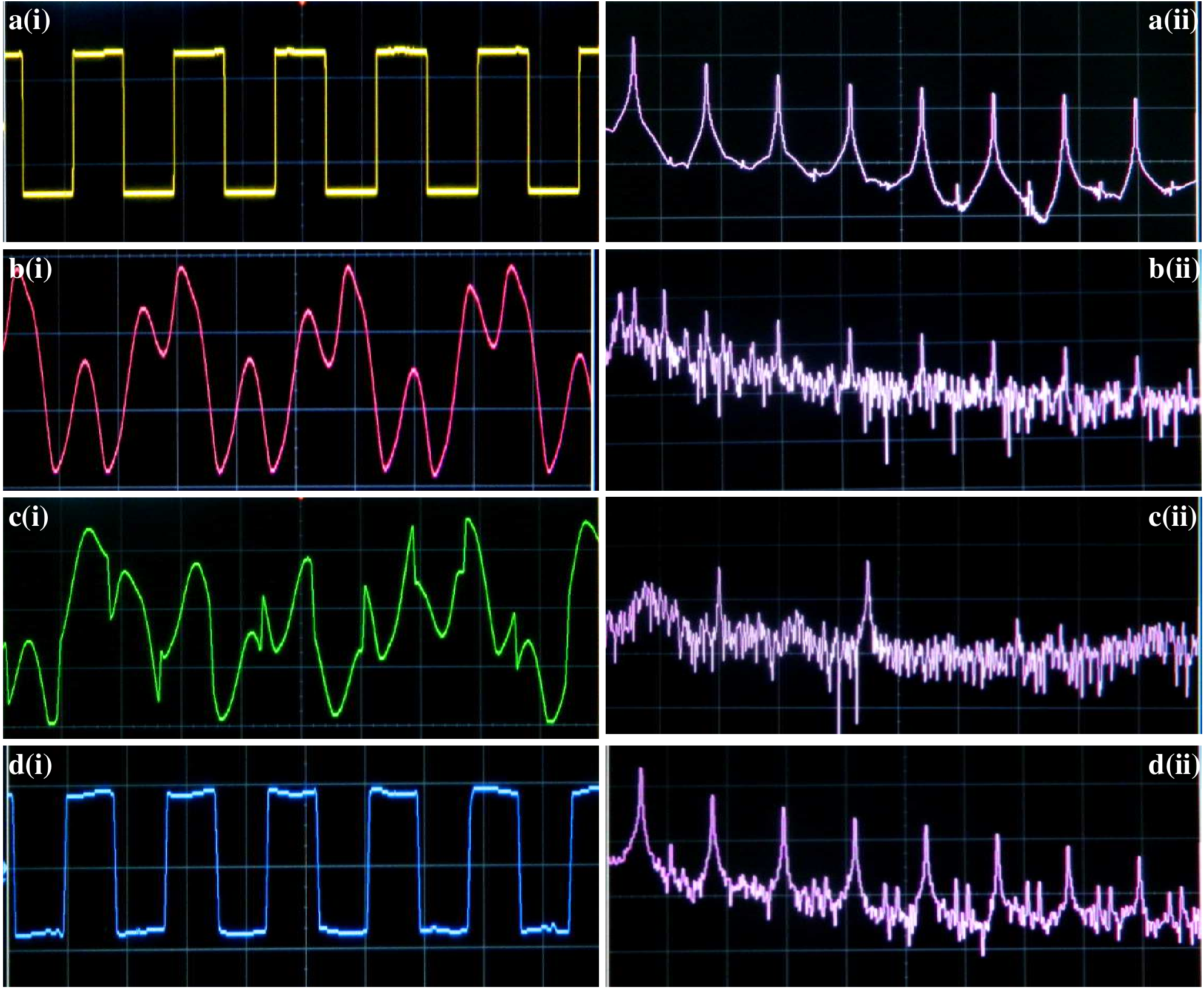}
\caption{Experimental implementation of the communication scheme observed using the unidirectionally coupled forced series {\emph{LCR}} circuits indicating the (i) time series and the corresponding (ii) power spectrum of the signals; (a) square wave information signal ($s(t)= F sin \omega t$, $F=0.5$ V, $\nu=6500$ Hz; (b) chaotic carrier signal $v(t)$; (c) the amplitude modulated chaotic transmitted signal $r(t) = v(t)+s(t)$ and (d) the recovered information signal $s^{'}(t)$ at the receiver.}
\label{fig:3}
\end{center}
\end{figure}
Figure \ref{fig:2}a(i)-d(i) represents the experimentally observed time series of the sine wave indicating the information signal {\emph{s(t)}}, the chaotic carrier wave {\emph{v(t)}} across the capacitor of the drive system, the amplitude modulated chaotic transmitted signal, $r(t)=v(t)+s(t)$ and the recovered information signal $s^{'}(t)$, respectively. The amplitude modulated wave shown in Fig. \ref{fig:2}b(i) indicates that the information signal has been completely masked by the carrier wave and retains its chaotic nature. Figures \ref{fig:2}a(ii)-d(ii) represent the corresponding power spectrum of the information signal {\emph{s(t)}}, chaotic carrier signal {\emph{v(t)}}, the amplitude modulated transmitted signal {\emph{r(t)}}  and the recovered information signal $s^{'}(t)$, respectively. As the power level of the information signal {\emph{s(t)}} is significantly lower than that of the chaotic carrier signal {\emph{v(t)}}, the frequency component of {\emph{s(t)}} is not detectable in Fig. \ref{fig:2}c(ii) due to the broadband nature of the modulated transmitted signal. The identical nature of the wave form and power spectra of the information signal and the recovered signal shown in Fig. \ref{fig:2}a(i), \ref{fig:2}a(ii) and Fig. \ref{fig:2}d(i), \ref{fig:2}d(ii) indicates the successful retrieval of the information signal at the receiver. This communication scheme has been further tested with another type of information bearing signal, the square wave, as shown in Fig. \ref{fig:3}. Figure \ref{fig:3}a(i)-d(i) represents the experimentally observed time waveform of the information bearing signal {\emph{s(t)}} (a square waveform), the chaotic carrier  waveform {\emph{v(t)}} of the drive system, the actual transmitted signal, $r(t)=v(t)+s(t)$ and the recovered signal $s^{'}(t)$, respectively. Figures \ref{fig:3}a(ii)-d(ii) represent the corresponding power spectrum of the information signal {\emph{s(t)}}, chaotic signal {\emph{v(t)}}, the actual transmitted signal {\emph{r(t)}}  and the recovered signal $s^{'}(t)$, respectively.

\section{Explicit analytical solutions}
\label{sec:3}

In this section, we present the explicit analytical solutions for the recovery of the information signal transmitted using a chaotic carrier signal. The analytical solutions are presented for the case of the information signal being a sine wave signal of the form $s(t)=f sin(\omega t)$. The amplitude of the information signal is so chosen that it is lesser than that of the chaotic carrier wave. After proper rescaling, the normalized state equations of the transmitter and the receiver systems given in Eqs. \ref{eqn:1} and \ref{eqn:2} can be written as\\
{\emph{Drive}} - Transmitter:\\
\begin{subequations}
\begin{eqnarray}
\dot x & = & y - g(x), \\
\dot y & = & - \sigma  y - \beta x + f_1 sin(\omega_1 t),
\end{eqnarray}
\label{eqn:5}
\end{subequations}
{\emph{Response}} - Receiver:
\begin{subequations}
\begin{eqnarray}
\dot {x^{'}} & = & y^{'} - g(x^{'}) + \epsilon (x+f sin(\omega t)-x^{'}), \\
\dot {y^{'}} & = & - \sigma  y^{'} - \beta x^{'} + f_2 sin( \omega_2 t),
\end{eqnarray}
\label{eqn:6}
\end{subequations}
where $ \beta = (C/LG^2)$, $ \nu = GR_s$, $ a  = G_a/G$,  $ b = G_b/G$, $f_{1,2} = (F_{1,2} \beta/B_p)$, $\omega_{1,2} = (\Omega_{1,2} C/G)$ and  $ G = 1/R$. The term $f sin(\omega t)$ represents the information signal and $g(x)$, $g(x^{'})$ represents the nonlinear element given as
\begin{equation}
g(x) =
\begin{cases}
bx+(a-b) & \text{if $x\ge 1$}\\
ax & \text{if $|x|\le 1$}\\
bx-(a-b) & \text{if $x\le -1$}
\end{cases}
\label{eqn:7}
\end{equation}
and
\begin{equation}
g(x^{'}) =
\begin{cases}
bx^{'}+(a-b) & \text{if $x^{'}\ge 1$}\\
ax^{'} & \text{if $|x^{'}|\le 1$}\\
bx^{'}-(a-b) & \text{if $x^{'}\le -1$}
\end{cases}
\label{eqn:8}
\end{equation}
When ($\epsilon=0$), the transmitter and the receiver systems are independent and the solutions are summarized as follows:\\
 
In the $D_0$ region, for complex roots, $y(t)$ and $x(t)$ are
\begin{subequations}
\begin{eqnarray}
y(t) &=& e^ {ut}(C_1 cos vt + C_2 sin vt) +E_1 + E_2 sin \omega_1 t \nonumber \\ 
&&+ E_3 cos \omega_1 t,\\
x(t) &=& \frac{1}{\beta}(-\sigma y - \dot{y} + f_1 sin \omega_1 t),
\end{eqnarray}
\label{eqn:9}
\end{subequations}
and for real roots
\begin{subequations}
\begin{eqnarray}
y(t) &=& C_1 e^ {m_1 t} + C_2 e^ {m_2 t} + E_1 + E_2 \sin \omega_1 t  \nonumber \\ 
&&+ E_3 \cos \omega_1 t,\\
x(t) &=& \frac{1}{\beta}(\dot{y}). 
\end{eqnarray}
\label{eqn:10}
\end{subequations}
In the $D_{\pm1}$ region, for complex roots
\begin{subequations}
\begin{eqnarray}
y(t) &=& e^ {ut}(C_3 cos vt + C_4 sin vt) +E_3 sin(\omega_1 t) \nonumber \\ 
&&+ E_4 cos(\omega_1 t) {\pm} \Delta,\\
x(t) &=& \frac{1}{\beta}(\dot{y}).
\end{eqnarray}
\label{eqn:11}
\end{subequations}
and for real roots
\begin{subequations}
\begin{eqnarray}
y(t) &=& C_3 e^ {m_3 t} + C_4 e^ {m_4 t} + E_3 \sin \omega_1 t  \nonumber \\ 
&&+ E_4 \cos \omega_1 t {\pm} \Delta,\\
x(t) &=& \frac{1}{\beta}(\dot{y}), 
\end{eqnarray}
\label{eqn:12}
\end{subequations}
The above solution is applicable for the receiver system operating with different initial conditions. For $\epsilon > 0$, the state equations of the transmitter and receiver systems given in Eqs. \ref{eqn:5} and \ref{eqn:6} gives rise to non-identical chaotic attractors owing to the presence of the information signal and the analytical solutions for retrieving the information signal can be deduced through suitable transformation of Eqs. \ref{eqn:5} and \ref{eqn:6}. The transformation is the formulation of a difference system in each of the coupled, identical piecewise-linear regions of the transmitter and receiver systems and hence, the 4-dimensional coupled system has been reduced to a 2-dimensional system. From Eqs. \ref{eqn:5} and \ref{eqn:6}, the difference system can be written as
\begin{subequations}
\begin{eqnarray}
\dot {x^{*}} & = & y^{*} - (g(x) -g(x^{'})) - \epsilon (x^{*}+f sin(\omega t)),\\
\dot {y^{*}} &= & -\sigma y^{*} - \beta x^{*} +f_1 sin(\omega_1 t) - f_2 sin(\omega_2 t).
\end{eqnarray}
\label{eqn:13}
\end{subequations}
where $x^{*}$=$(x-x^{'})$, $y^{*}$=$(y-y^{'})$ and $g(x) -g(x^{'})=g(x^{*})$ take the values $a{x^{*}}$ or $b{x^{*}}$ depending on the region of operation. From the relations $r(t)=s(t)+x(t)$ and $x^{*}(t)=x(t)-x^{'}(t)$, we get
\begin{equation}
s(t)=r(t)-x(t)=r(t)-x^{*}(t)-x^{'}(t) \approx -x^{*}(t),
\label{eqn:14}
\end{equation}
Hence, when the amplitude modulated signal $r(t)$ is nearly equal to the chaotic signal of the receiver system $x^{'}(t)$, {\emph{i.e.}} $r(t) \approx x^{'}(t)$, the state variable $x^{*}(t)$ indicates a measure of the information signal retrieved at the receiver, given as
\begin{equation}
s^{'}(t)=-x^{*}(t),
\label{eqn:15}
\end{equation}
The above idea of the recovered signal $s^{'}(t)$ can be confirmed by obtaining analytical solutions for the condition $\epsilon > 0$. Now, we find the solution of  the state variables $x^{*}(t), y^{*}(t)$ in the regions $D^{*}_{0}$ and $D^{*}_{\pm1}$ of the difference system. Analytical solutions of this kind has been studied recently for synchronization in a number of systems \cite{Sivaganesh2015,Sivaganesh2017,Sivaganesh2018,Sivaganesh2018b}. Hence, we summarize the solution for the state equations of the difference system as follows.

\subsection{ \bf $D^{*}_0~region$}

For real and distinct roots, the state variables are given as 
\begin{eqnarray}
y^{*}(t) = && C_1 e^ {m_1 t} + C_2 e^ {m_2 t} + E_1 sin(\omega_1 t) + E_2 cos(\omega_1 t) + E_3 sin(\omega_2 t) \nonumber \\
&&  + E_4 cos(\omega_2 t) + E_5 sin(\omega t) + E_6 cos(\omega t)
\label{eqn:16}
\end{eqnarray}

\begin{equation}
x^{*}(t) = (\frac{1}{\beta})(\dot{y^{*}} - \sigma y^{*} + f_1 sin(\omega_1 t) - f_2 sin(\omega_2 t))
\label{eqn:17}
\end{equation}

The constants are
\begin{subequations}
\begin{eqnarray}
E_1  &=&  \frac {f_1  {\omega_1} ^2 (A-a-\epsilon) + f_1 B(a+\epsilon)}{A^2 {\omega_1} ^2 + (B-{\omega_1} ^2)^2}  \\
E_2  &=&  \frac {f_1  \omega_1 ((B-{\omega_1} ^2)-A(a+\epsilon ))}{A^2 {\omega_1} ^2 + (B-{\omega_1} ^2)^2}  \\
E_3  &=&   -\frac {f_2  {\omega_2} ^2 (A-a-\epsilon) + f_2 B(a+\epsilon)}{A^2 {\omega_2} ^2 + (B-{\omega_2} ^2)^2}  \\
E_4  &=&  -\frac {f_2  \omega_2 ((B-{\omega_2} ^2)-A(a+\epsilon ))}{A^2 {\omega_2} ^2 + (B-{\omega_2} ^2)^2} \\
E_5  &=&   \frac {\beta \epsilon f (B-{\omega_2} ^2)}{A^2 {\omega} ^2 + (B-{\omega} ^2)^2}  \\
E_6  &=&   -\frac {A \omega \beta \epsilon f}{A^2 {\omega} ^2 + (B-{\omega} ^2)^2} 
\end{eqnarray}
\label{eqn:18}
\end{subequations}
\begin{eqnarray}
C_1 =  &&\frac{e^ {- m_1 t_0}} {m_1 - m_2} \{ (-\sigma{ y^{*}_0}-\beta{ x^{*}_0}-m_2{ y^{*}_0}) - (\omega_1 E_1 - m_2 E_2) cos \omega_1 t_0 \nonumber\\
&& + ( f_1+ \omega_1 E_2 + m_2 E_1) sin \omega_1 t_0  -(\omega_2 E_3 - m_2 E_4) cos \omega_2 t_0 \nonumber \\
&&  + (\omega_2 E_4 + m_2 E_3 - f_2) sin \omega_2 t_0 - (\omega E_5 - m_2 E_6) cos \omega t_0 \nonumber \\
&& + (\omega E_6 + m_2 E_5) sin \omega t_0 \} \nonumber
\end{eqnarray}
\begin{eqnarray}
C_2 =  && \frac{e^ {- m_2 t_0}} {m_2 - m_1} \{ (-\sigma{ y^{*}_0}-\beta{ x^{*}_0}-m_1{ y^{*}_0}) - (\omega_1 E_1 - m_1 E_2) cos \omega_1 t_0 \nonumber \\
&& + ( f_1+ \omega_1 E_2 + m_1 E_1) sin \omega_1 t_0 -(\omega_2 E_3 - m_1 E_4) cos \omega_2 t_0 \nonumber \\
&& + (\omega_2 E_4 + m_1 E_3 - f_2) sin \omega_2 t_0 - (\omega E_5 - m_1 E_6) cos \omega t_0 \nonumber \\
&& + (\omega E_6 + m_1 E_5) sin \omega t_0 \} \nonumber
\end{eqnarray}

For complex roots, 
\begin{eqnarray}
y^{*}(t) =  && e^ {ut} (C_1 cosvt+ C_2 sinvt) + E_1 sin \omega_1 t + E_2 cos \omega_1 t + E_3 sin \omega_2 t  \nonumber \\
&& + E_4 cos \omega_2 t + E_5 sin(\omega t) + E_6 cos(\omega t)
\label{eqn:19}
\end{eqnarray}

\begin{equation}
x^{*}(t) = (\frac{1}{\beta})(-\dot{y^{*}} - \sigma y^{*}+ f_1 sin(\omega_1 t) - f_2 sin(\omega_2 t))
\label{eqn:20}
\end{equation}

and the  constants $C_1$ and $C_2$ are
\begin{eqnarray}
C_1 = && \frac{e^ {- u t_0}} {v} \{((\sigma { y^{*}_0} +\beta { x^{*}_0}+u { y^{*}_0})sinvt_0 + v { y^{*}_0}) cosvt_0 \nonumber \\
&& +((\omega_1 E_1 - u E_2) sinvt_0 - vE_2 cosvt_0)cos \omega_1 t_0 \nonumber \\
&& - ((f_1+\omega_1 E_2 +u E_1) sinvt_0+v E_1 cosvt_0) sin \omega_1 t_0 \nonumber \\
&& + ((\omega_2 E_3 - u E_4) sinvt_0 - vE_4 cosvt_0)cos \omega_2 t_0  \nonumber \\
&&  - ((\omega_2 E_4 + u E_3-f_2) sinvt_0 + vE_3 cosvt_0) sin \omega_2 t_0  \nonumber \\
&& + ((\omega E_5 - u E_6) sinvt_0 - vE_6 cosvt_0)cos \omega t_0  \nonumber \\
&& - ((\omega E_6 + u E_5) sinvt_0 + vE_5 cosvt_0)sin \omega t_0 \}  \nonumber
\end{eqnarray}
\begin{eqnarray}	 
C_2 = && \frac{e^ {- u t_0}} {v} \{((-\sigma { y^{*}_0} -\beta { x^{*}_0} - u { y^{*}_0})cos vt_0 + v { y^{*}_0}) sin vt_0 \nonumber \\
&& - ((\omega_1 E_1 - u E_2) cos vt_0 + vE_2 sin vt_0)cos \omega_1 t_0 \nonumber \\
&&  + ((f_1+\omega_1 E_2 +u E_1) cos vt_0 - v E_1 sin vt_0) sin \omega_1 t_0 \nonumber \\
&& - ((\omega_2 E_3 - u E_4) cos vt_0 + vE_4 sin vt_0)cos \omega_2 t_0  \nonumber \\
&& + ((\omega_2 E_4 + u E_3 - f_2) cos vt_0 - vE_3 sin vt_0) sin \omega_2 t_0 \nonumber \\
&& - ((\omega E_5 - u E_6) cos vt_0 + vE_6 sin vt_0)cos \omega t_0  \nonumber \\
&& + ((\omega E_6 + u E_5) cos vt_0 - vE_5 sin vt_0)sin \omega t_0 \}  \nonumber
\end{eqnarray}

\subsection{ \bf $D^{*}_{\pm1}~region$}

The state variables for real and distinct roots are
\begin{eqnarray}
y^{*}(t) = && C_3 e^ {m_3 t} + C_4 e^ {m_4 t} + E_7 sin(\omega_1 t) + E_8 cos(\omega_1 t) + E_9 sin(\omega_2 t) \nonumber \\
&&  + E_{10} cos(\omega_2 t) + E_{11} sin(\omega t) + E_{12} cos(\omega t)
\label{eqn:21}
\end{eqnarray}
\begin{equation}
x^{*}(t) = (\frac{1}{\beta})(-\dot{y^{*}} - \sigma y^{*}+ f_1 sin(\omega_1 t) - f_2 sin(\omega_2 t))
\label{eqn:22}
\end{equation}
and for complex roots
\begin{eqnarray}
y^{*}(t)  =&&  e^ {ut} (C_3 cosvt+ C_4 sinvt)+ E_7 sin(\omega_1 t) + E_8 cos(\omega_1 t) + E_9 sin(\omega_2 t) \nonumber \\
&&  + E_{10} cos(\omega_2 t) + E_{11} sin(\omega t) + E_{12} cos(\omega t)
\label{eqn:23}
\end{eqnarray}
\begin{equation}
x^{*}(t) = (\frac{1}{\beta})(-\dot{y^{*}} - \sigma y^{*}+ f_1 sin(\omega_1 t) - f_2 sin(\omega_2 t))
\label{eqn:24}
\end{equation}
The state variables of the difference system $x^{*}(t)$ and $y^{*}(t)$ obtained at every of instant is used to recover the information signal using Eq. \ref{eqn:15}.\\
\begin{figure}
\begin{center}
\includegraphics[scale=0.5]{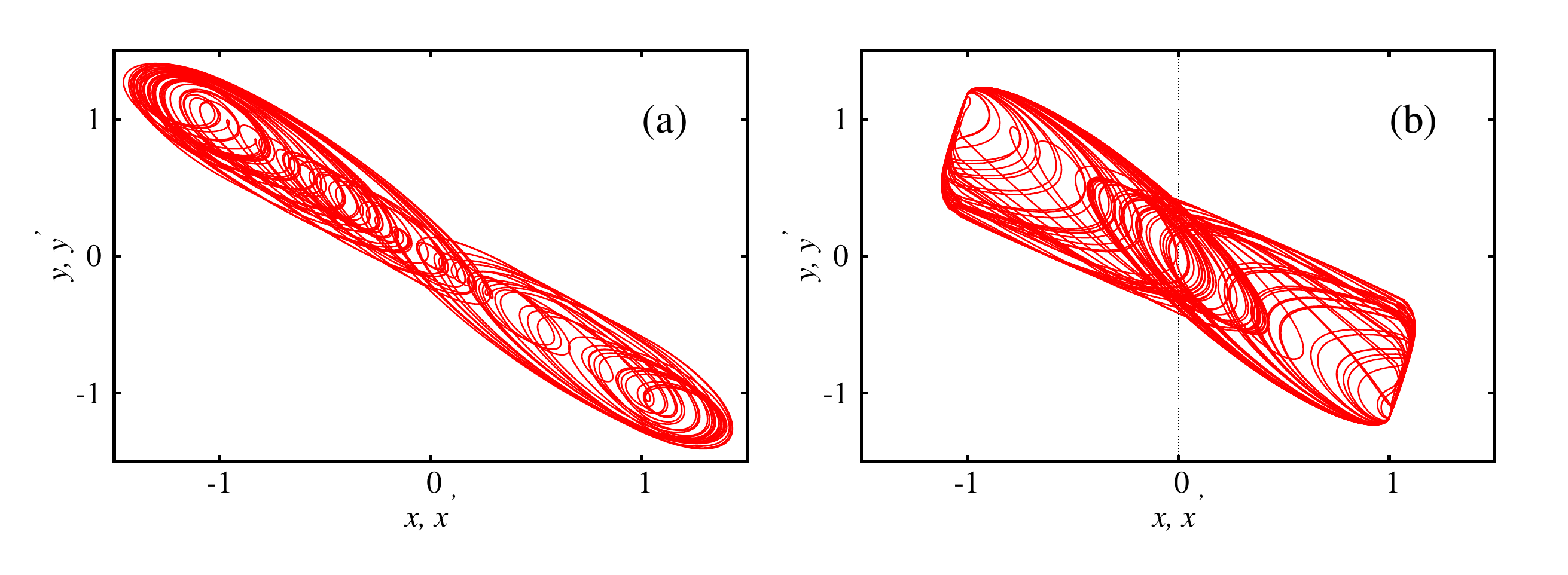}
\caption{Double band chaotic attractors of the transmitter and receiver systems in the $x-y$ and $x^{'}-y^{'}$ phase planes obtained using the analytical solutions. Chaotic attractor of the (a) {\emph{MLC}} circuit and (b) the series {\emph{LCR}} circuit with a {\emph{SNE}}.}
\label{fig:4}
\end{center}
\end{figure}
\begin{figure}
\begin{center}
\includegraphics[scale=0.33]{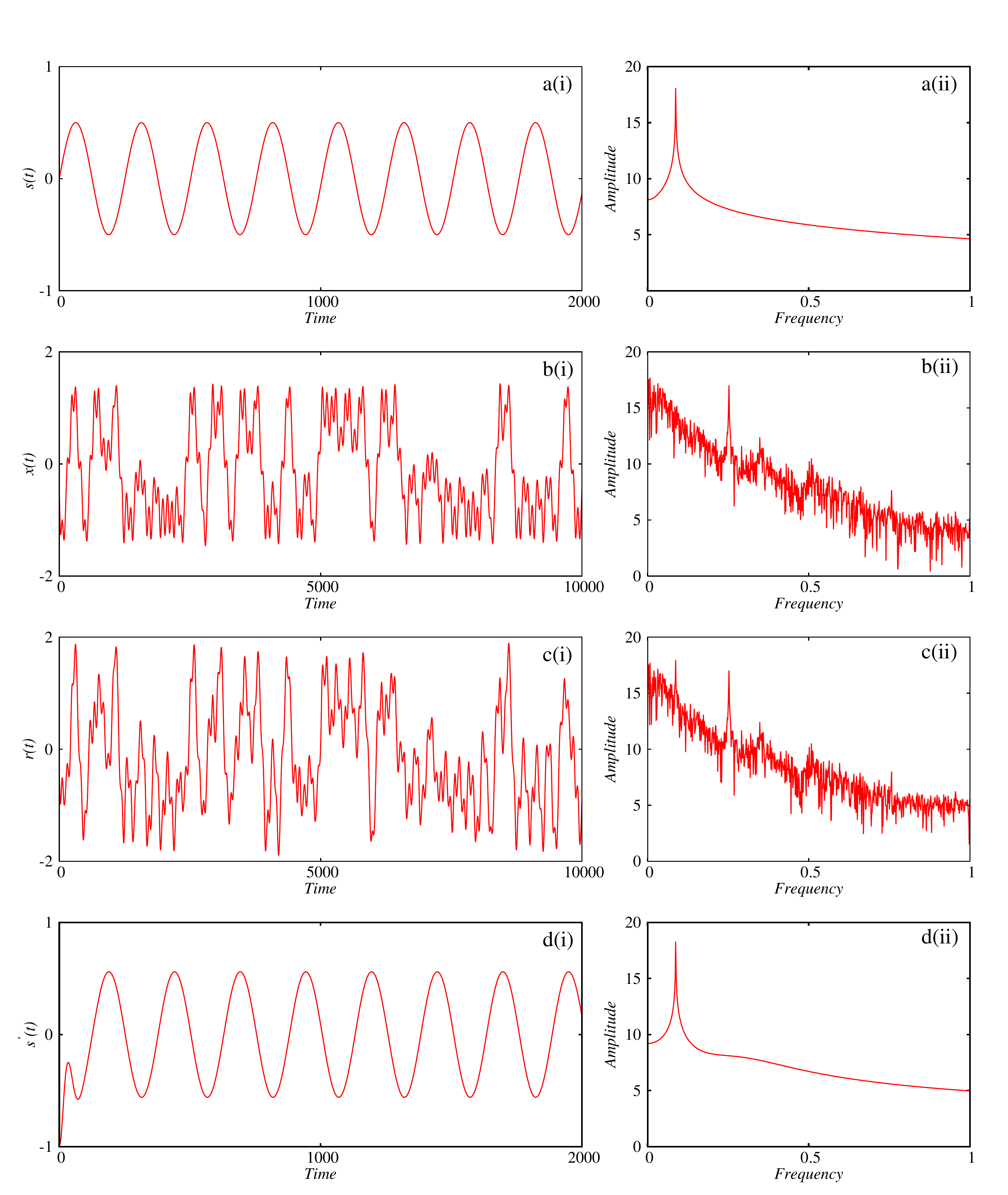}
\caption{{\emph{MLC}} circuit. Analytical results obtained for $\epsilon=0.85$ indicating the (i) time series and (ii) power spectrum; (a) original information signal ($s(t)= f sin \omega t$, $F=0.5, \nu=0.25$ Hz); (b) chaotic carrier signal $x(t)$; (c) amplitude modulated transmitted signal $r(t) = x(t)+s(t)$ and (d) the difference signal $x^{*}(t)$ representing the recovered information signal $s^{'}(t)$.}
\label{fig:5}
\end{center}
\end{figure}
\begin{figure}
\begin{center}
\includegraphics[scale=0.33]{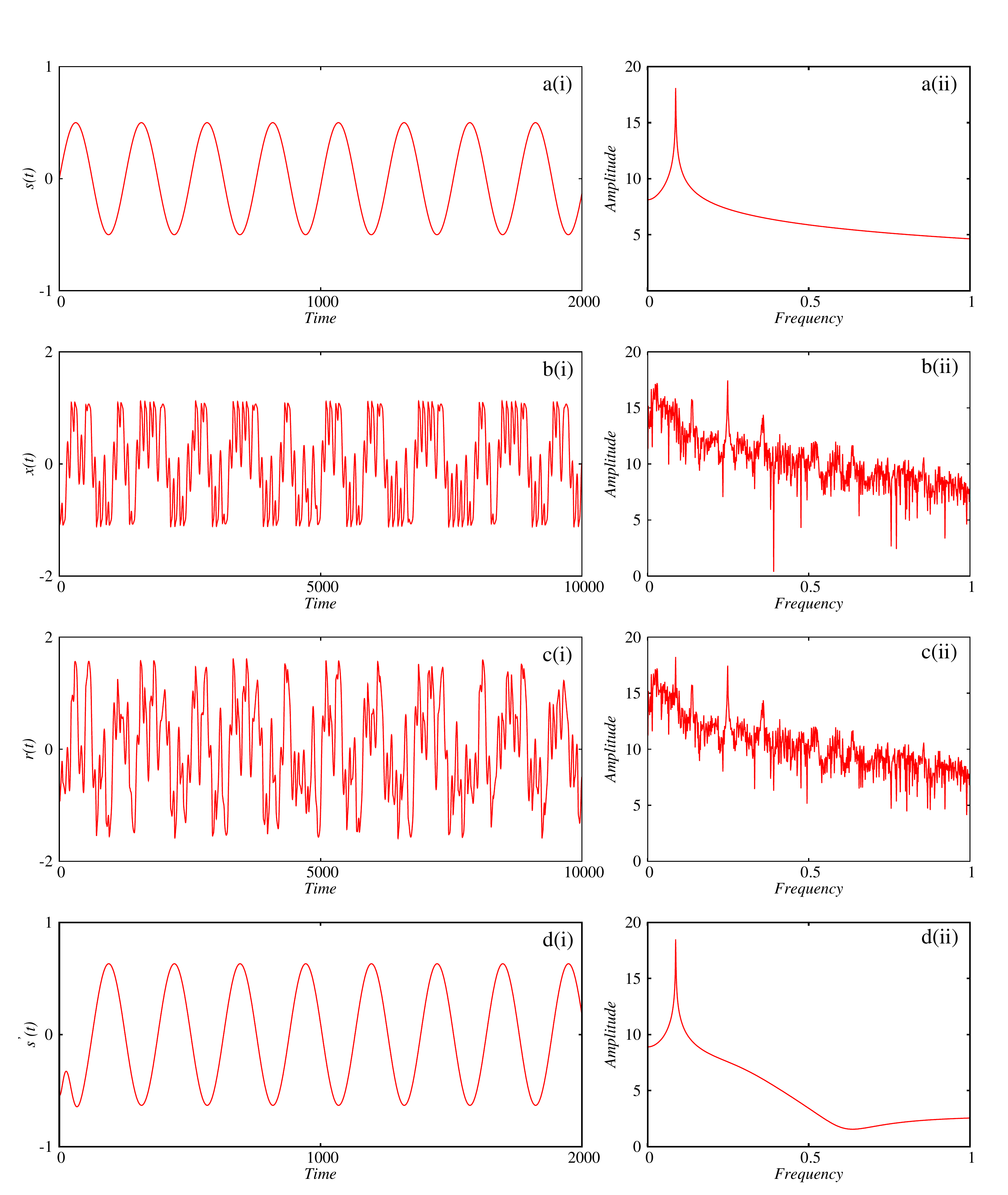}
\caption{Series {\emph{LCR}} circuit with {\emph{SNE}}. Analytical results obtained for $\epsilon=1$ indicating the (i) time series (ii) power spectrum; (a) original information signal ($s(t)= f sin \omega t$, $F=0.5, \nu=0.25$ Hz); (b) chaotic carrier signal $x(t)$; (c) amplitude modulated transmitted signal $r(t) = x(t)+s(t)$ and (d) the difference signal $x^{*}(t)$ representing the recovered information signal $s^{'}(t)$.}
\label{fig:6}
\end{center}
\end{figure}
\begin{figure}
\begin{center}
\includegraphics[scale=1]{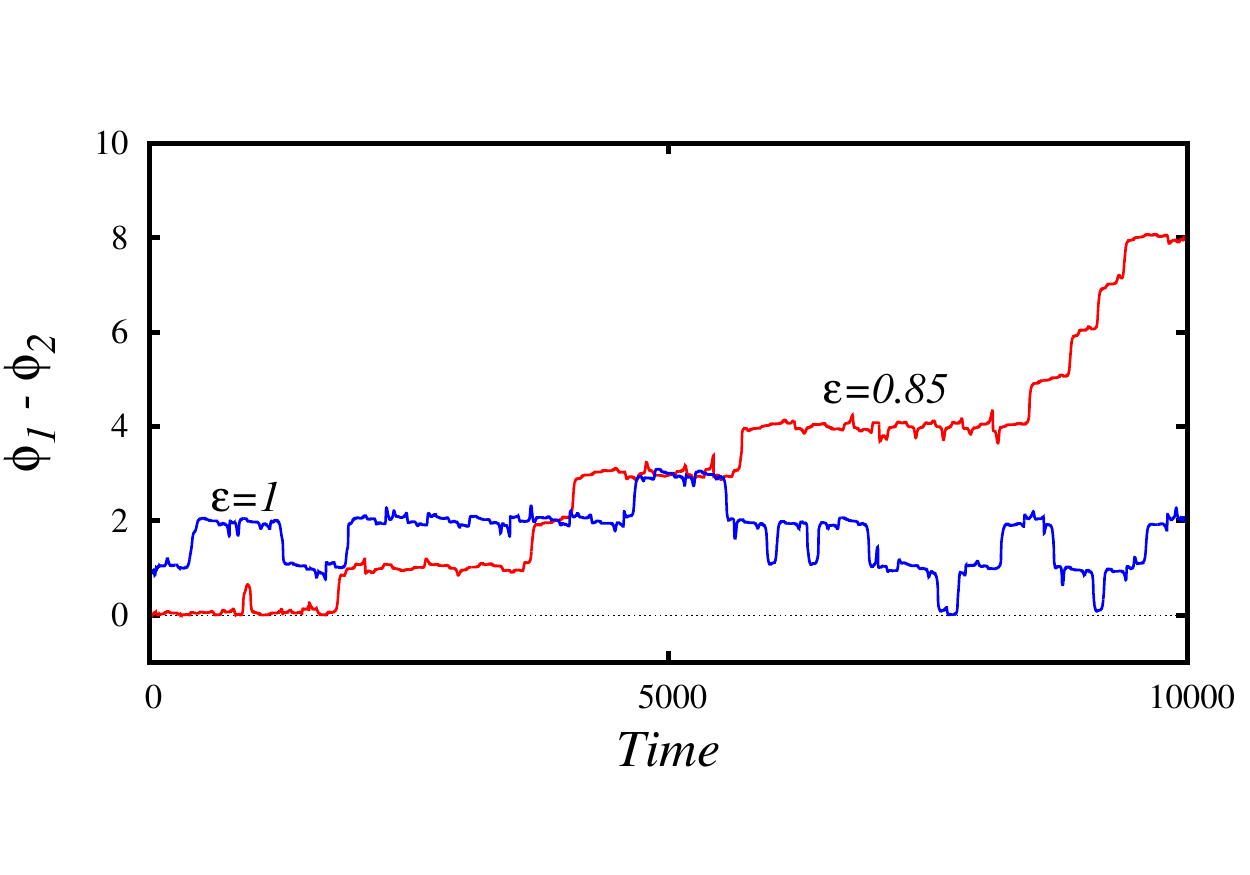}
\caption{Time-series of the phase difference $(\phi_1 - \phi_2)$ of the transmitter and receiver systems of unidirectionally coupled {\emph{MLC}} circuits (red) and {\emph{SNE}} circuits (blue) indicating imperfect locking of the phases for the coupling strengths $\epsilon=0.85$ and $\epsilon=1$, respectively.}
\label{fig:7}
\end{center}
\end{figure}
The analytical solutions presented above is used to study the transmission of the information signals through chaotic masking. Two types of chaotic systems that provide the carrier signals are considered for the present study. In the first case, we consider the chaotic signal of the {\emph{Murali-Lakshmanan-Chua (MLC)}} circuit and in the second case the series {\emph{LCR}} circuit with a {\emph{SNE}} is considered. In both the cases, the chaotic signals corresponds to the double-band chaotic attractor states. The chaotic attractors of the transmitter and receiver systems obtained from the analytical solutions given in Eqs. \ref{eqn:9}-\ref{eqn:12} in the $x-y$ and $x^{'}-y^{'}$ planes corresponding to the MLC circuit and the {\emph{SNE}} circuit are shown in Fig. \ref{fig:4}(a) and \ref{fig:4}(b), respectively. \\

The MLC circuit is a second-order, non-autonomous chaotic circuit system having the {\emph{Chua's diode}} as a nonlinear element \cite{Murali1994a} with the system parameters $a=-1.02$, $b=-0.55$, $\beta=1$, $\nu=0.015$, $f_{1,2}=0.14$, $\omega_{1,2}=0.72$. The amplitude and frequency of the sinusoidal information signal are fixed as $f=0.5$ and$\omega=0.25$, respectively. The recovery of the information signal obtained from the explicit analytical solutions with the MLC circuit acting as the chaotic carrier is shown in Fig. \ref{fig:5}. Figures \ref{fig:5}a(i), b(i) and c(i) represents the analytically observed time series of the sinusoidal information signal {\emph{s(t)}}, the chaotic carrier signal {\emph{x(t)}} corresponding to the transmitter system and the amplitude modulated transmitted signal, $r(t)=x(t)+s(t)$, respectively for the coupling strength $\epsilon=0$. The power spectra corresponding to the time series given in Fig. \ref{fig:5}a(i)-c(i) is shown in Fig. \ref{fig:5}a(ii)-c(ii). The power spectra of the amplitude modulated chaotic transmitted wave embedding the information signal shown in Fig. \ref{fig:5}c(ii) indicates a broad band nature resembling that of the chaotic carrier wave shown in Fig. \ref{fig:5}b(ii) and hence the information signal is undetectable and securely masked by the chaotic carrier wave. The time series of the state variable $x^{*}(t)$ of the difference system representing the information signal $s^{'}(t)$ as given by Eq. \ref{eqn:15} and its corresponding power spectrum obtained for the coupling strength $\epsilon=0.85$ are shown in Fig. \ref{fig:5}d(i) and d(ii), respectively. The power spectra of the information signal $s^{'}(t)$ recovered at the receiver shown in Fig. \ref{fig:5}d(ii) indicates a single tone frequency nature resembling the original information signal shown in Fig. \ref{fig:5}a(ii). Hence, the solution of the difference system $x^{*}(t)$ qualitatively represents the recovered signal $s^{'}(t)$ as given in Eq. \ref{eqn:15}. The applicability of the generalized analytical solution presented above is studied with another circuit system namely, the series {\emph{LCR}} circuit with a {\emph{simplified nonlinear element}}.\\

The series {\emph{LCR}} circuit with a {\emph{SNE}} introduced by {\emph{Arulgnanam et al}} \cite{Arulgnanam2009} has the system parameter values $a=-1.148$, $b=5.125$, $\beta=0.9865$, $\nu=0$, $f_{1,2}=0.31$, $\omega_{1,2}=0.7084$. The amplitude and frequency of the sinusoidal information signal is fixed as $f=0.5$ and $\omega=0.25$, respectively. The secure transmission of this information signal using the chaotic carrier signal of the {\emph{SNE}} system is as shown in Fig. \ref{fig:6}. Figures \ref{fig:6}a(i), b(i) and c(i) represents the analytically observed time series of the sinusoidal information signal {\emph{s(t)}}, the chaotic carrier signal {\emph{x(t)}} corresponding to the transmitter system and the amplitude modulated transmitted signal, $r(t)=x(t)+s(t)$, respectively for the coupling strength $\epsilon=0$. The power spectra corresponding to the time series given in Fig. \ref{fig:6}a(i)-c(i) is shown in Fig. \ref{fig:6}a(ii)-c(ii). Fig. \ref{fig:6}c(i) and c(ii) representing the time series and power spectra of the amplitude modulated chaotic transmitted wave indicates that the information signal is undetectable and securely masked by the chaotic carrier wave. The time series of the state variable $x^{*}(t)$ of the difference system representing the information signal $s^{'}(t)$ as given by Eq. \ref{eqn:15} and its corresponding power spectrum obtained for the coupling strength $\epsilon=1$ are shown in Fig. \ref{fig:6}d(i) and d(ii), respectively. The power spectra of the information signal $s^{'}(t)$ recovered at the receiver shown in Fig. \ref{fig:6}d(ii) indicates a single tone frequency nature resembling the original information signal shown in Fig. \ref{fig:6}a(ii). Hence, it can confirmed that the solution of the difference system $x^{*}(t)$ qualitatively represents the recovered signal $s^{'}(t)$ as given in Eq. \ref{eqn:15}. The recovered signal shown in Fig. \ref{fig:6}d(i) indicates enhancement in the amplitude of the signal as compared to the original signal shown in \ref{fig:6}a(i).\\

The nature of synchronization phenomena observed in the coupled systems can be analyzed to reveal the mechanism involved on the recovery of the information signal. From Fig. \ref{fig:1}, it could be identified that the transmitter and the receiver systems are non-identical owing to the inclusion of the information signal at the transmitter. Recent studies on the synchronization behavior of coupled non-identical second-order, non-autonomous systems indicates the entrainment of the phases of the coupled systems with the increase in coupling strength \cite{Sivaganesh2018}. Figure \ref{fig:7} indicates the imperfect locking of the phases of the coupled MLC (red) and {\emph{simplified nonlinear element}} (blue) circuit systems for the coupling strengths $\epsilon=0.85$ and $\epsilon=1$, respectively. Hence, the signals corresponding to the state variables of the coupled systems undergo imperfect locking of phases and the transmitted and recovered information signals always have a phase difference as shown in Fig. \ref{fig:7}. From the above discussions, it is observed that the solution of the difference system $(x^{*}(t))$ is a qualitative measure of the information signal $(s^{'}(t))$ recovered at the receiver system as given by Eq. \ref{eqn:15}. This behavior has been confirmed in two types of chaotic systems discussed above and the nature of the synchronization phenomena observed in the coupled systems during the signal recovery is identified.

\begin{figure}
\begin{center}
\includegraphics[scale=0.33]{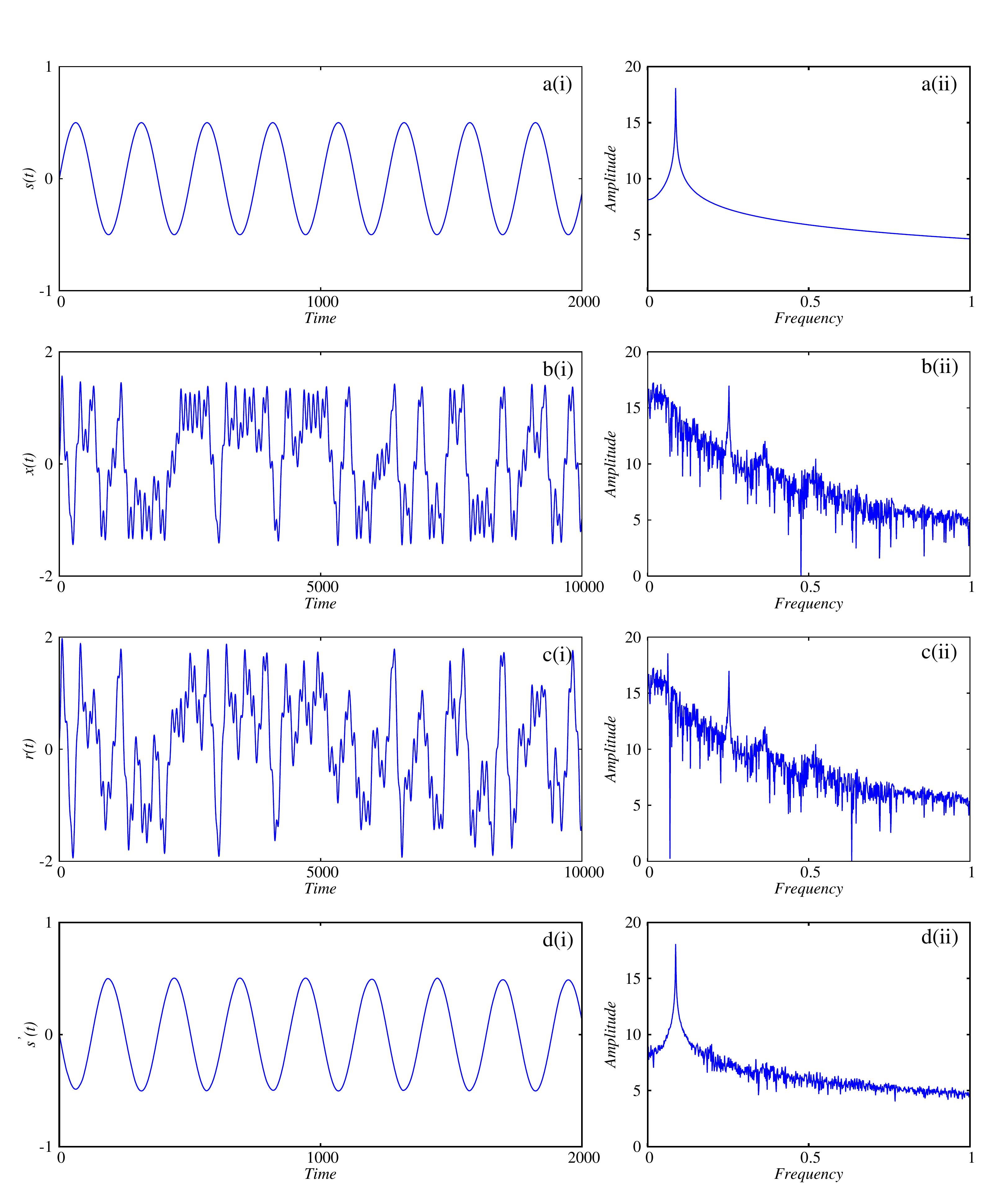}
\caption{{\emph{MLC}} circuit. Numerical results obtained for $\epsilon=12$ indicating the (i) time series and (ii) power spectrum; (a) original information signal ($s(t)= f sin \omega t$, $F=0.5, \nu=0.25$ Hz); (b) chaotic carrier signal $x(t)$; (c) amplitude modulated transmitted signal $r(t) = x(t)+s(t)$ and (d) the recovered information signal $s^{'}(t)$.}
\label{fig:8}
\end{center}
\end{figure}
\begin{figure}
\begin{center}
\includegraphics[scale=0.33]{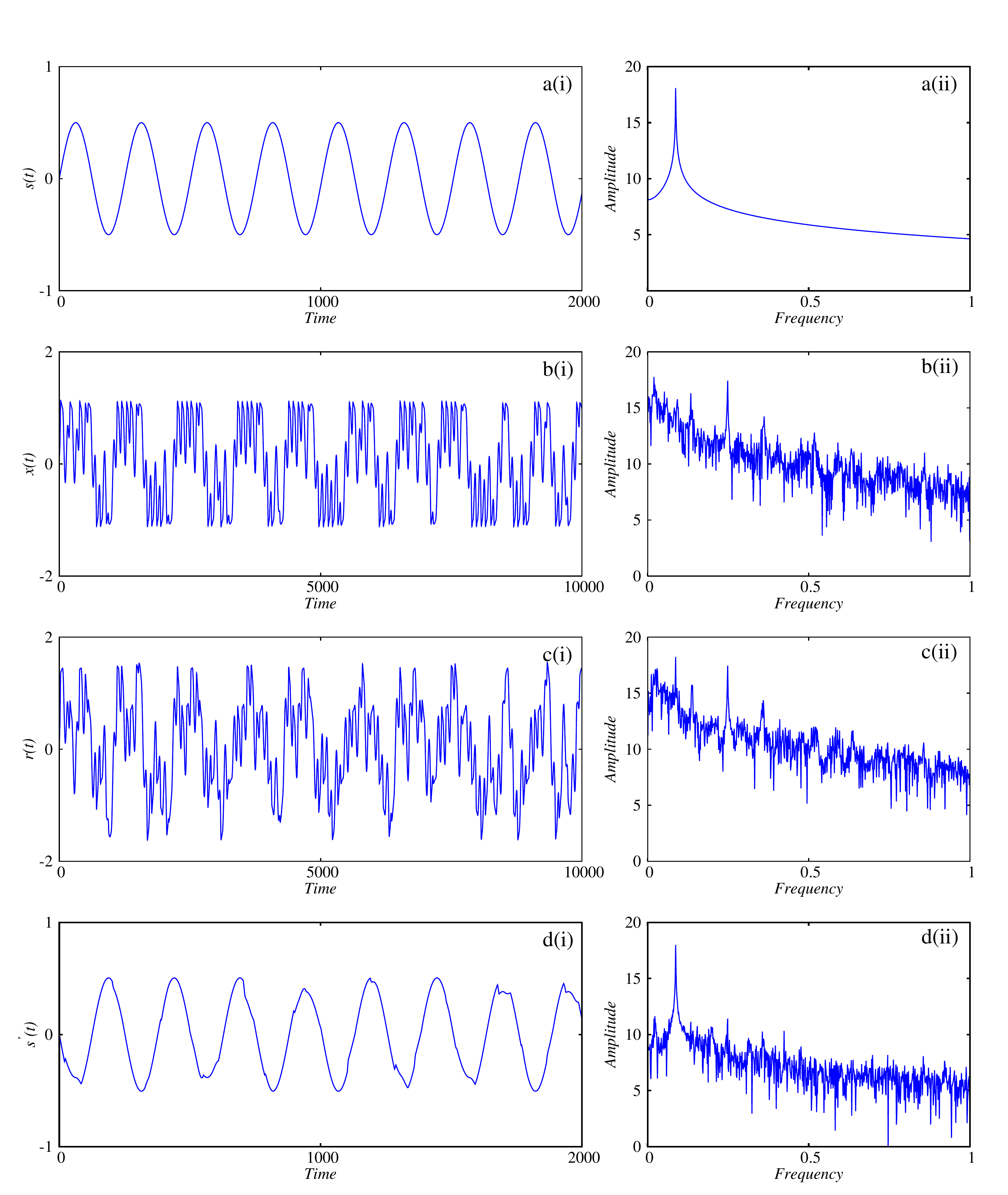}
\caption{Series {\emph{LCR}} circuit with {\emph{SNE}}. Numerical results obtained for $\epsilon=20$ indicating the (i) time series (ii) power spectrum; (a) original information signal ($s(t)= f sin \omega t$, $F=0.5, \nu=0.25$ Hz); (b) chaotic carrier signal $x(t)$; (c) amplitude modulated transmitted signal $r(t) = x(t)+s(t)$ and (d) the recovered information signal $s^{'}(t)$.}
\label{fig:9}
\end{center}
\end{figure}
\begin{figure}
\begin{center}
\includegraphics[scale=0.5]{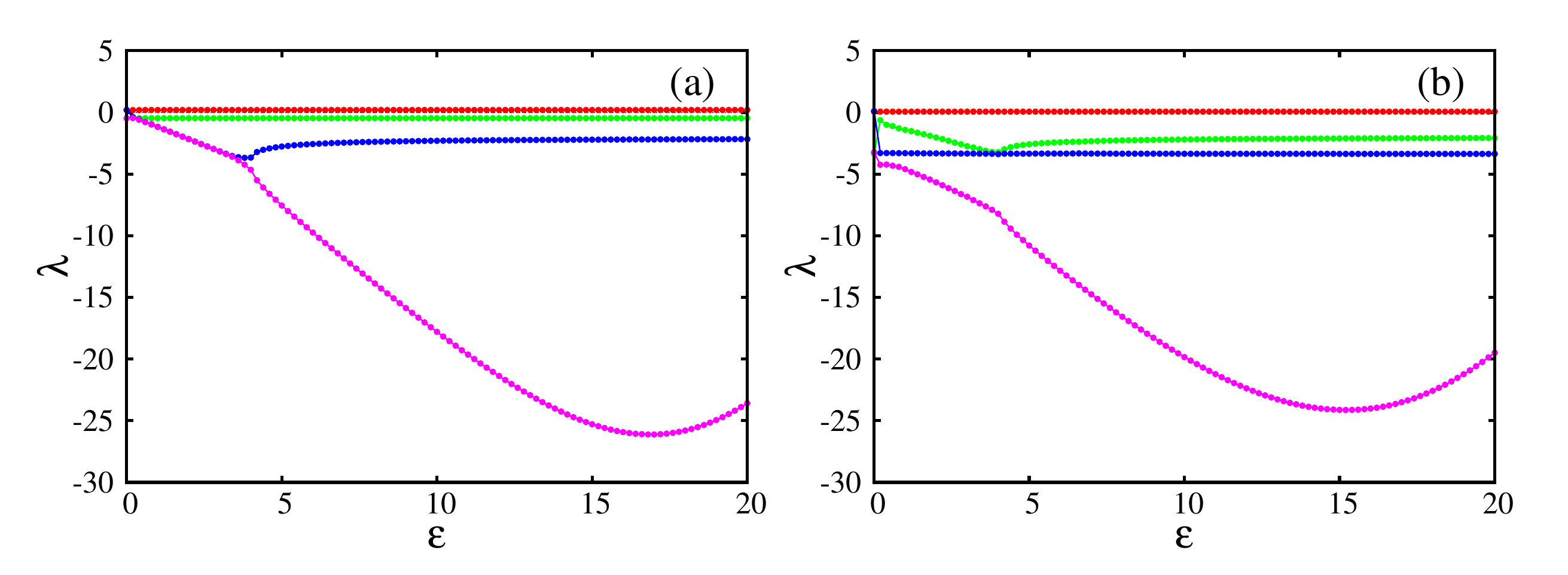}
\caption{The four largest Lyapunov exponents of the unidirectionally coupled chaotic systems corresponding to (a) {\emph{MLC}} circuits and (b) series {\emph{LCR}} circuits with {\emph{SNE}}.}
\label{fig:10}
\end{center}
\end{figure}

\section{Numerical results}
\label{sec:4}

In this section, we present the numerical simulation results for signal transmission achieved through synchronization of chaotic systems. The normalized state equations of the coupled transmitter and receiver system given in Eqs. \ref{eqn:5} and \ref{eqn:6} are simulated for two different system parameters corresponding to the the MLC and {\emph{SNE}} circuits to validate the analytical results. Figure \ref{fig:8} shows the numerical results obtained for the system parameters $a=-1.02$, $b=-0.55$, $\beta=1$, $\nu=0.015$, $f_{1,2}=0.14$, $\omega_{1,2}=0.72$ corresponding to the MLC circuit system. Figures \ref{fig:8}a(i), b(i) and c(i) represents the numerically observed time series of the sinusoidal information signal {\emph{s(t)}}, the chaotic carrier signal {\emph{x(t)}} and the amplitude modulated transmitted signal, $r(t)=x(t)+s(t)$, respectively for the coupling strength $\epsilon=0$. The power spectra corresponding to the time series of the signals given in Fig. \ref{fig:8}a(i)-c(i) are shown in Fig. \ref{fig:8}a(ii)-c(ii). The undetectable nature of the information signal owing to chaotic masking is represented by the amplitude modulated transmitted wave and its corresponding power spectrum shown in Fig. \ref{fig:8}c(i) and c(ii), respectively. The time series of the recovered information signal $s^{'}(t)$ and its corresponding power spectrum obtained for the coupling strength $\epsilon=12$ are shown in Fig. \ref{fig:8}d(i) and d(ii), respectively. The power spectra of the information signal $s^{'}(t)$ recovered at the receiver shown in Fig. \ref{fig:8}d(ii) indicates a single tone frequency nature resembling the original information signal shown in Fig. \ref{fig:8}a(ii). It has to be noted that the recovered information signal $s^{'}(t)$ presented in Fig. \ref{fig:8}d(i) is actually the difference signal $x(t)-x^{'}(t)$ i.e. $s^{'}(t)=x(t)-x^{'}(t)$.\\

 The numerical results showing the recovery of information signal observed in coupled series {\emph{LCR}} circuits with {\emph{SNE}} for the system parameters $a=-1.148$, $b=5.125$, $\beta=0.9865$, $\nu=0$, $f_{1,2}=0.31$, $\omega_{1,2}=0.7084$ is shown in Fig. \ref{fig:9}. Figures \ref{fig:9}a(i), b(i) and c(i) represents the numerically observed time series of the sinusoidal information signal {\emph{s(t)}}, the chaotic carrier signal {\emph{x(t)}} and the amplitude modulated transmitted signal, $r(t)=x(t)+s(t)$, respectively for the coupling strength $\epsilon=0$. The power spectra corresponding to the time series of the signals given in Fig. \ref{fig:9}a(i)-c(i) are shown in Fig. \ref{fig:9}a(ii)-c(ii). The undetectable nature of the information signal owing to chaotic masking is represented by the amplitude modulated transmitted wave and its corresponding power spectrum shown in Fig. \ref{fig:9}c(i) and c(ii), respectively. The time series of the recovered information signal $s^{'}(t)$ and its corresponding power spectrum obtained for the coupling strength $\epsilon=20$ are shown in Fig. \ref{fig:9}d(i) and d(ii), respectively. The power spectra of the information signal $s^{'}(t)$ recovered at the receiver shown in Fig. \ref{fig:9}d(ii) indicates a single tone frequency nature resembling the original information signal shown in Fig. \ref{fig:9}a(ii). The broadband nature of the recovered signal observed in the power spectra shown in Fig. \ref{fig:8}d(ii) and \ref{fig:9}d(ii) are due to numerical artifacts. The largest Lyapunov exponents of the coupled system given by Eqs. \ref{eqn:5} and \ref{eqn:6} indicating the synchronized nature for the two types of system parameters are presented in Fig. \ref{fig:10}. 
 
\section{Conclusion}
\label{sec:5}
We have reported in this paper, an explicit generalized analytical solution for a simple communication scheme involved in the transmission and recovery of an information signal. The waveform of two simple chaotic systems have been used as the carrier to mask the information signal. The analytical results leads to the following important conclusions. The state variable of the difference system indicates a measure of the information signal recovered at the receiver. The transmitter and the receiver systems does not synchronize completely and undergo imperfect locking of their phases at higher values of coupling strength. The amplitude of the recovered signal has been enhanced during its retrieval at the receiver. The numerical results presented validates the claim that the recovered signal is obtained from the difference of the coupled state variables. The analytical and numerical results are confirmed through electronic circuit experimental results. The transmission and recovery of information signal using a simple communication scheme studied through explicit analytical solutions is reported in the literature for the first time.

\bibliography{mybibfile_1}

\end{document}